\documentclass{article}

\usepackage{PRIMEarxiv}

\usepackage[utf8]{inputenc} 
\usepackage[T1]{fontenc}    
\usepackage{hyperref}       
\usepackage{url}            
\usepackage{booktabs}       
\usepackage{amsfonts}       
\usepackage{nicefrac}       
\usepackage{microtype}      
\usepackage{lipsum}
\usepackage{fancyhdr}       
\usepackage{graphicx}       
\graphicspath{{media/}}     
\usepackage{physics}
\usepackage{lineno}
\usepackage{amssymb}
\title{$N$-Band Photonic Hopf Insulators Based on 2D Microring Lattices
}

\author{
 Bo Leng, Vien Van \\
 Department of Electrical and Computer Engineering \\
 University of Alberta\\
  Edmonton, Alberta, T6G 2W3, Canada\\
  \texttt{\{Bo Leng\}bleng@ualberta.ca} \\
}

\begin{document}
\maketitle

\begin{abstract}
Hopf insulators are topological insulators whose topological behavior arises from the nontrivial mapping from a 3D sphere to a 2D sphere, known as the Hopf map.  The Hopf map, typically encountered in the study of spinor and skyrmion systems, is classified topologically by an integer invariant called the Hopf index.  Here we show that due to the periodic circulation of light inside each microring, a 2D lattice of microring resonators can emulate an $N$-band photonic Hopf insulator with nontrivial Hopf index.  In particular, we show by numerical computation and direct analytical proof that the $N$-band Hopf index of the microring lattice is identical to its winding number.  The result shows that the Hopf index is an alternative topological invariant for classifying 2D microring photonic lattices and establishes a correspondence between the Hopf insulator phase and the anomalous Floquet insulator phase of the lattice.  More generally, our work shows that 2D microring lattices can provide a versatile nanophotonic platform for studying non-Abelian topological photonic systems.
\end{abstract}

\keywords{Hopf insulator \and Topological photonics}

\section*{Introduction}
Topological insulators are band insulators whose energy bands or band gaps are characterized by global invariants that remain unchanged in the presence of disorders.  In addition to their importance in the study of solid-state systems, these materials can potentially have applications in realizing electronic \cite{taskin2012manifestation,hsieh2009tunable,bahari2017nonreciprocal} and photonic devices \cite{bandres2018topological,wang2019topological,dai2022topologically,zimmerling2022broadband}  that are robust to imperfections.  The "ten-fold" classification has successfully provided a systematic way to classify a large number of topological insulators based on lattice symmetry \cite{kitaev2009periodic}.  However, there also exist topological systems beyond the ten-fold classification which exhibit topologically protected edge states even though the relevant topological invariants (e.g. Chern numbers) of all the energy bands are trivial.  One prominent example is the anomalous Floquet insulator (AFI) phase of periodically driven systems \cite{rudner2013anomalous, afzal2020realization,afzal2018topological,kitagawa2010topological,pasek2014network,nathan2015topological,leykam2016anomalous,maczewsky2017observation,mukherjee2017experimental}, whose band gaps can have nontrivial winding number even when the Chern invariants of all the quasi-energy bands vanish.  Another topological insulator which exists beyond the ten-fold classification is the Hopf insulator, first investigated in static 3D magnetic topological insulator with two energy bands \cite{moore2008topological}.  When all relevant weak invariants vanish, the system can still exhibit nontrivial behavior and is classified by the Hopf index, which is related to the linking number of the Hopf map $S^{3}\rightarrow S^2$ \cite{pontrjagin1941,makhlin1995topology,faddeev1997stable,wilczek1983linking} typically encountered in spinor and skyrmion systems \cite{kent2021creation,sutcliffe2017skyrmion,rybakov2019stable,voinescu2020hopf}.  
Interestingly, it was later shown \cite{unal2019hopf} that 
this Hopf index exactly corresponds to the winding number of a 2D+1 Floquet insulator with two quasi-energy bands, thus establishing a connection between the Hopf phase and the AFI phase.  
The two-band Hopf insulator was generalized to the multi-band case in \cite{lapierre2021n}, where it was shown that a 3D system with $N > 2$ distinct bands and vanishing weak invariants can be classified by an $N$-band Hopf index associated with the third homotopy group $\pi_3(SU(N))$.  It should be stressed that a defining feature of Hopf insulators is that they do not possess any symmetry of the "ten-fold" classification so that their nontrivial topological behavior arises solely from the Hopf map.

The Hopf invariant has previously been used to characterize dynamical quenching processes of 2D cold atom topological systems, enabling the Chern number of the system under equilibrium to be experimentally measured \cite{wang2017scheme,tarnowski2019measuring}.
In the photonics domain, there have so far been no reported realizations of Hopf insulators, although AFI insulators have been demonstrated using coupled waveguide arrays\cite{leykam2016anomalous,mukherjee2017experimental,maczewsky2017observation} and microring lattices \cite{afzal2020realization,dai2022topologically,zimmerling2022broadband}.  However, not all AFIs are governed by Hamiltonians that are homotopic to the Hopf map and are thus Hopf insulators.   In this paper, we show that 2D square lattices of coupled microrings can emulate an $N$-band Hopf insulator with $N = 3$ and $N = 4$.  The direction of light circulation in each microring serves as the third periodic dimension so that our system is a map from $T^2\times S^1 \rightarrow SU(N)$.  In particular, we show by numerical simulations as well as by direct analytical proof that the $N$-band Hopf index of the microring lattice is identical to the winding number of each band gap and can thus provide a complete topological classification of the insulator.  Both $N = 3$ and $N = 4$ band microring lattices have been experimentally demonstrated before as AFIs \cite{afzal2020realization,zimmerling2022broadband,dai2022topologically}.  Here we show that these systems can also be regarded as photonic Hopf insulators, thereby providing further insight into the nontrivial topological behaviors of 2D microring lattices and demonstrating their versatility as a nanophotonic platform for studying non-Abelian topological photonic systems.


\section*{Hopf insulator and its multi-band generalization}
We begin by first giving a brief review of Hopf insulator and its multi-band generalization.  Given any 2D insulator with two energy bands separated by a band gap, we can express its Hamiltonian as $H(k_x,k_y) = U(k_x,k_y) \sigma_3 U^{\dagger} (k_x,k_y)$, where $\sigma_3 = \mathrm{diag}[1,-1]$ is a Pauli matrix and $U(k_x,k_y) \in SU(2)$ is a $2 \times 2$ unitary transformation matrix which flattens the bands to have constant energy spectra of $\pm1$. We can regard $H(k_x,k_y)$ as a map of each point $(k_x, k_y)$ in the 2-torus ($T^2$) Brillouin zone (BZ) to the quotient group $SU(2)/U(1)$, where $U(1)$ accounts for the fact that only the relative phase between the two Bloch eigenstates is fixed \cite{chiu2016classification}.  Since the group $SU(2)/U(1)$ is isomorphic to a 2-sphere ($S^2$), $H(k_x,k_y)$ is homotopic to the map $T^2 \rightarrow S^2$. The topological invariant classifying this map is the Chern number, $C_z =\int_{BZ} dk_x dk_y B_z$,
where $B_z$ is the $z$-component of the Berry curvature.  This corresponds to the well-known classification of 2D Chern insulators.  
For a 3D insulator with the same two-band constraint, the Hamiltonian $H(\vec{k})$, $\vec{k} = (k_x,k_y,k_z)$, becomes a map from the 3-torus ($T^3$) BZ to the group $SU(2)/U(1)$, or $T^3 \rightarrow S^2$.  A 3-torus has three submanifolds, each being a cross-section $T^2$ perpendicular to the axis $k_x$, $k_y$, or $k_z$. As in the 2D case, each submanifold is classified by a Chern number corresponding to a component $\mu = x, y, z$ of the Berry curvature, 
$
    C_\mu =\int_{BZ} dk_\rho dk_\lambda B_\mu,
$
where $dk_\rho dk_\lambda$ describes a cross-section of the $T^3$ BZ perpendicular to the $\mu$ direction \cite{deng2013hopf}. When all three invariants vanish, $H(\vec{k})$ becomes homotopic to the Hopf map, $S^3 \rightarrow S^2$, which can be nontrivial since it is classified by the third homotopy group $\pi_3(S^2) = \mathbb{Z}$ \cite{govindarajan1990inequivalent}.  Thus, even though the Chern numbers of the submanifolds are zero, the insulator can still be topologically nontrivial.  Such a 3D two-band topological insulator is called a Hopf insulator \cite{moore2008topological,deng2013hopf}.  
Under the assumption of vanishing weak invariants ($C_x = C_y = C_z = 0$), the topological invariant of this map is given by the Hopf index, $n_H = 2P_3^1 \in \mathbb{Z}$, where $P_3^n$ is the Abelian third Chern-Simons form of band $n$,
\begin{equation}
    P_3^n=\int_{BZ} \frac{d^3\vec{k}}{8\pi^2} \Vec{A}_n\cdot \nabla_{\vec{k}} \times \Vec{A}_n \label{eq:3csA}
\end{equation}
with $\vec{A}_n=i\bra{\psi_n}\nabla_{\vec{k}}\ket{\psi_n}$ being the Berry connection associated with the $n^{\mathrm{th}}$ Bloch state $|\psi_n(\vec{k})\rangle$. 

The two-band Hopf insulator can be generalized to multi-band systems in 3D, which are called $N$-band Hopf insulators \cite{lapierre2021n}. The Hamiltonian of an insulator with $N$ distinct flattened bands can be expressed as $H(\vec{k})=U\text{diag}[1,\ldots,N]U^\dagger$,
 where $U(\vec{k}) \in SU(N)$ is an $N \times N$ unitary matrix.  The Hamiltonian in this case is a map from the 3-torus BZ to the quotient group $SU(N)/U(1)^{N-1}$ \cite{lapierre2021n}.  Under the same assumption of vanishing weak invariants, this map is classified by the third homotopy group $\pi_3(SU(N)/U(1)^{N-1}) \cong \pi_3(SU(N))$.  The topological invariant is also given by the Hopf index but in this case computed from the non-Abelian third Chern-Simons form,
\begin{equation}
    n_H =\int_{BZ} \frac{d^3\vec{k}}{8\pi^2} \text{Tr}\bigg\{ \vec{\hat{A}}\cdot \nabla_{\vec{k}} \times \vec{\hat{A}} -\frac{2i}{3} \vec{\hat{A}} \cdot \vec{\hat{A}} \times \vec{\hat{A}} \bigg\} \label{eq:HopfIndex}
\end{equation}
Here $\vec{\hat{A}}$ is the matrix-valued connection, $\vec{\hat{A}} = \hat{A}_{k_x}\vec{\textbf{k}}_x + \hat{A}_{k_y}\vec{\textbf{k}}_y + \hat{A}_{k_z}\vec{\textbf{k}}_z$, where the $\vec{\textbf{k}}_{\mu}$-component $\hat{A}_{k_{\mu}}$ is an $N \times N$ matrix with elements $(\hat{A}_{k_{\mu}})_{mn}=i\expval{\psi_m|\partial_{k_{\mu}}\psi_n }$. 

Since the Hamiltonian of a 3D insulator can be emulated by a 2D periodically-driven system with the substitution $k_z\rightarrow 2\pi t/T$, where $T$ is the driving period, there is a correspondence between a Hopf insulator and the AFI phase of a Floquet insulator in 2D+1 dimensions \cite{unal2019hopf,lapierre2021n}.  In particular, the AFI phase exists when the Chern numbers of all the quasi-energy bands of a Floquet insulator vanish and is classified by the third winding number of the evolution operator $\mathcal{U}$, $W[\mathcal{U}]$. For a 2D+1 Floquet insulator, it can be shown that the winding number is identical to the Hopf invariant, $W = n_H$ \cite{unal2019hopf,lapierre2021n}.

\section*{Photonic Hopf Insulator Based on 2D Microring Lattice}
The photonic Hopf insulator we consider is a 2D square lattice of evanescently-coupled microring resonators with identical resonance frequencies, as shown in Fig. 1(a). Each unit cell consists of four microrings labeled $\{A, B, C, D\}$ characterized by two unequal coupling angles: $\theta_a$ between microring $A$ and its neighbors, and $\theta_b$ between microring $D$ and its neighbors.  The coupling angles correspond to power coupling coefficients $\kappa_{(a,b)}^2=\sin^2\theta_{(a,b)}$ between adjacent resonators.  We assume that each microring in the lattice supports only a clockwise or counterclockwise propagating mode.  As light circulates around each microring, it couples periodically to its neighbors, with a period $L$ equal to the microring circumference.  The direction of light propagation in each microring, denoted by $z$, thus takes the role of time in our periodically-driven system.  Using waveguide coupled mode theory, the evolution of the system can be described by a Schrodinger-like equation (see Supplementary Materials for derivation),
\begin{equation}
    -i\frac{\partial}{\partial z}|\psi(\vec{k},z)\rangle=[\beta I + H_{FB}(\vec{k},z)]|\psi(\vec{k},z)\rangle 
    \label{eq;sh}
\end{equation}
where $I$ is the $4\times4$ identity matrix, $\beta$ is the propagation constant of the microring waveguides, $\vec{k} = (k_x, k_y)$ is the crystal momentum, and $|\psi(\vec{k},z)\rangle = [\psi_A, \psi_B, \psi_C, \psi_D]^{\mathrm{T}}$ is the wave function representing the fields in the four microrings in each unit cell. The Floquet-Bloch Hamiltonian $H_{FB}$ describes a hopping sequence of four coupling steps, $ H_{FB}(\vec{k},z)=\sum_{j=1}^4 H_j(\vec{k})$, where $H_j$ in step $j$ is given by
\begin{equation}
    \begin{array}{ll}
     H_1(\vec{k}) = K\otimes \sigma_1, & 0 \leq z < L/4\\
     H_2(\vec{k}) = \sigma_1 \otimes K, & L/4 \leq z < L/2\\
     H_3(\vec{k}) = K \otimes (\sigma_1 \cos{k_x} + \sigma_2 \sin{k_x}), & L/2 \leq z < 3L/4\\
     H_4(\vec{k}) = (\sigma_1 \cos{k_y} + \sigma_2 \sin{k_y}) \otimes K, & 3L/4 \leq z < L
     \end{array}
\end{equation}
with $K = \mathrm{diag}[\theta_a L/4, \theta_b L/4]$ and $\sigma_i$ being the Pauli matrices.
The Hamiltonian is periodic in $z$ with periodicity $L$, $H_{FB}(\vec{k},z)=H_{FB}(\vec{k},z+L)$.  Neglecting the dynamic phase term ($e^{i\beta z}$), the evolution operator of the lattice is given by $\mathcal{U}(\vec{k},z) = \mathcal{T} e^{i\int_0^z H_{FB}(\vec{k},z')dz'}$, where $\mathcal{T}$ is the time-order operator.  
Over one period $L$, the system evolution is captured by the Floquet operator $U_F(\vec{k}) = \mathcal{U}(\vec{k},L)$, whose eigenstates are the Floquet modes $\Phi_n(\vec{k})$,
 \begin{equation}
     U_F(\vec{k})|\Phi_n(\vec{k})\rangle=e^{i\epsilon_n(\vec{k})L}|\Phi_n(\vec{k})\rangle\label{eq:eig}
 \end{equation}
 with associated quasi-energy bands $\epsilon_n(\vec{k})$. These bands can be regarded as the energy bands of a 2D static system with an effective Hamiltonian $H_{eff}(\vec{k})$ defined such that $U_F(\vec{k}) = e^{i H_{eff}(\vec{k})L}$.
 Over each Floquet-Brillouin zone, the microring lattice has $N = 4$ quasi-energy bands separated by 3 gaps, labeled I, II and III, as shown in Fig. 1(b) for the case ($\theta_a,\theta_b) = (0.48\pi,0.05\pi)$.  Band gap II is centered at quasi-energy $\epsilon =\pi/L$ and is always open except for $\theta_a = \theta_b$ or $\theta^2_a+\theta^2_b\approx \pi/8$, while band gaps I and III are symmetric about $\epsilon= 0$ and are only open for certain range of ($\theta_a, \theta_b$) values (red and blue areas in Fig. 2(a)). When $\theta_b$ is set to 0, which corresponds to removal of microring $D$ from each unit cell, the two bands around quasi-energy $\epsilon = 0$ collapse to form a single flat band.  In this case the microring lattice forms a Lieb lattice \cite{mukherjee2015observation} with $N = 3$ quasi-energy bands, with a flat band pinned at $\epsilon = 0$, as shown in Fig. 1(c).  For each quasi-energy band, we can compute the Chern number from the Floquet state as\cite{rudner2013anomalous}
\begin{equation}
    C_z^n=\frac{1}{2\pi i}\int_{BZ}\text{Tr}\bigg\{ P_n\left[ \partial_{k_x}P_n,\partial_{k_y}P_n\right]\bigg\} dk_x dk_y
    \label{eq:Cn}
\end{equation}
where $P_n(\vec{k})=|\Phi_n(\vec{k})\rangle \langle \Phi_n(\vec{k})|$ is the projection operator onto the Floquet state $\Phi_n(\vec{k})$. 
In general, the Chern number is zero for quasi-energy bands $n = 1$ and 4 but nontrivial for bands $n = 2$ and 3 over the range of $(\theta_a, \theta_b)$ values shown by the red areas in Fig. 2(a).  In this region the microring lattice exhibits Chern insulator behavior in band gaps I and III \cite{afzal2018topological}.

\begin{figure}[t]
    \centering
    \includegraphics[scale=0.5]{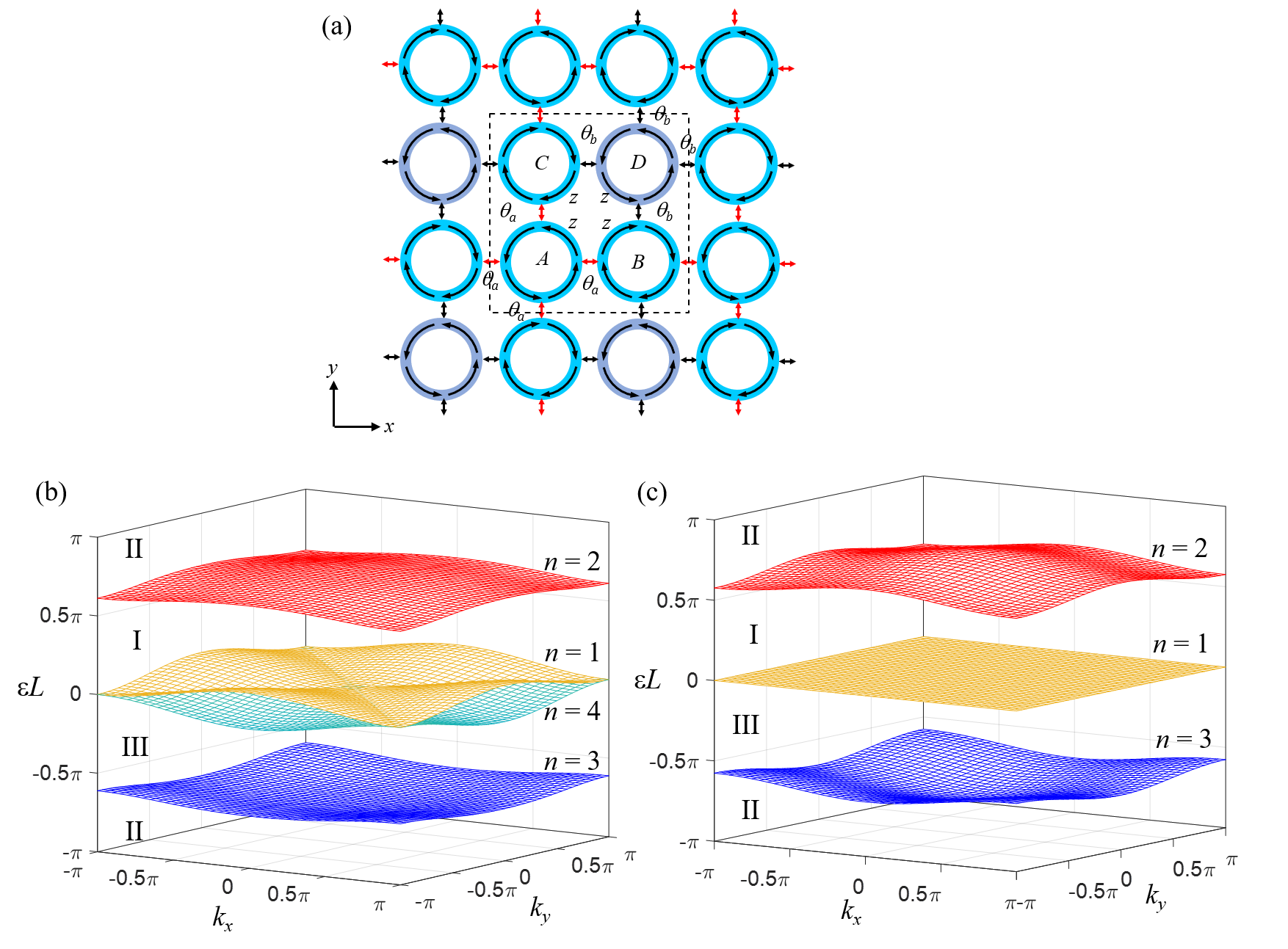}
    \caption{(a) Schematic of a 2D microring lattice characterized by dissimilar coupling angles $\theta_a \neq \theta_b$ in each 4-ring unit cell $\{A, B, C, D\}$.  In each microring, light propagates in either the clockwise or counterclockwise direction, which is denoted by $z$. (b) and (c) Quasi-energy bands and band gaps of (b) $N = 4$ band lattice with $\theta_a = 0.48\pi, \theta_b = 0.05 \pi$ and (c) $N=3$ band lattice with $\theta_a = 0.46\pi, \theta_b = 0$. }
    \label{fig:Fig1}
\end{figure}

Due to the periodic circulation of light in each microring, the Floquet-Bloch Hamiltonian $H_{FB}(\vec{k},z)$ of the $N$-band microring lattice can be regarded as a map from $T^2 \times S^1$ to the group $SU(N)/U(1)^{(N-1)}$.
The Chern number computed in Eq.(\ref{eq:Cn}) classifies only the 2-torus sub-manifold $k_x-k_y$ in each Floquet-Brillouin zone and thus does not completely capture the full topological properties of the lattice.  In particular, when the Chern numbers of all the bands vanish, $H_{FB}$ is homotopic to the map $S^2 \times S^1 \rightarrow SU(N)/U(1)^{(N-1)} \cong \pi_3(SU(N))$, which is classified by the $N$-band Hopf index in Eq.(\ref{eq:HopfIndex}).
Computation of the Hopf index requires us to first construct the connection matrix $\vec{\hat{A}}$, which is a non-Abelian gauge potential since the gauge transformation matrices $U \in SU(N)$ acting on the states of the $N$-band microring lattice are non-commutative. For our Floquet system, we can construct the connection matrix $\vec{\hat{A}}$ using the $z$-evolved Floquet state $\Psi_{n,\xi}(\vec{k},z)$ defined as
\begin{equation}
    |\Psi_{n,\xi}(\vec{k},z)\rangle=\mathcal{U}_\xi(\vec{k},z)|\Phi_n(\vec{k})\rangle 
    \label{eq:propagatingmode}
\end{equation}
where $\mathcal{U}_\xi(\vec{k},z) =\mathcal{U}(\vec{k},z) e^{-iH_{eff,\xi}(\vec{k})z}$ is the periodized evolution operator with the property $\mathcal{U}_\xi(\vec{k},0)=\mathcal{U}_\xi(\vec{k},L)=I$.  Here the effective Hamiltonian $H_{eff,\xi}(\vec{k}) = -(i/L)ln(U_F(\vec{k}))$ is defined for the band gap at energy $\xi$ by choosing its eigenvalues to be between $\xi$ and $\xi+2\pi/L$ \cite{rudner2013anomalous}.  The $z$-evolved Floquet states are periodic, $|\Psi_{n,\xi}(\vec{k},L)\rangle = |\Psi_{n,\xi}(\vec{k},0)\rangle$, and orthogonal $\langle\Psi_{m,\xi}(\vec{k},z)|\Psi_{n,\xi}(\vec{k},z)\rangle = \delta_{mn}$, since the periodic evolution operator preserves the orthogonality of the initial Floquet states during evolution.  The elements of the connection matrix are computed as $(\hat{A}_{\mu})_{mn}=i\langle \Psi_{m,\xi}(\vec{k},z)| \partial_{\mu} \Psi_{n,\xi}(\vec{k},z)\rangle$, where $\mu \in \{k_x,k_y,z\}$. The Hopf index can then be written explicitly in terms of the $z$-evolved Floquet states as
\begin{equation}\label{eq:nh_numberical}
\begin{split}
    n_H &= -\frac{1}{8\pi^2}\int_0^L dz \int_{BZ}  dk_x dk_y (F_1 + F_2)\\
    &
    F_1 = \sum_{n=1}^{N}\sum_{m=1}^N\epsilon_{i j k}\langle \Psi_{m,\xi} | \partial_i \Psi_{n,\xi} \rangle \langle \partial_j\Psi_{n,\xi} | \partial_k \Psi_{m,\xi} \rangle\\
    &F_2 = \sum_{l=1}^N \sum_{m=1}^N \sum_{n=1}^N \epsilon_{i j k}\langle\Psi_{l,\xi}| \partial_i\Psi_{m,\xi} \rangle \langle\Psi_{m,\xi}| \partial_j\Psi_{n,\xi}\rangle
    \langle\Psi_{n,\xi} | \partial_k\Psi_{l,\xi} \rangle 
\end{split}
\end{equation}
where $\epsilon_{ijk}$ is the Levi-Civita symbol representing summation over $k_x,k_y,z$. From its definition, the $N$-band Hopf index is a band gap invariant which is summed over the topological structures of all the bands anchored at quasi-energy $\xi$.

We computed the Hopf invariant for each open band gap of the microring lattice as a function of the coupling angles $\theta_a$ and $\theta_b$.  The results are shown in Fig. 2(b). For band gaps I and III, the Hopf invariant is equal to 1 so the lattice is always nontrivial as long as these gaps are open. 
For band gap II, the Hopf index transitions from 0 to 1 when the coupling angles satisfy the approximate relation $\theta_a^2+\theta_b^2 \gtrapprox \pi^2/8$. 
For the $N = 3$ band lattice, which corresponds to the $\theta_b = 0$ (or $\theta_a = 0$) axis of the map, the Hopf index is 1 in band gap I for all values of coupling angle $\theta_a$ while for band gap II, it transitions from 0 to 1 for $\theta_a > \pi/\sqrt{8}$.  The Hopf insulator phase is defined as having nontrivial Hopf invariant when the Chern numbers of all the quasi-energy bands vanish.  This region is highlighted by the red lines in Fig. 2(b) showing the existence of the nontrivial Hopf phase for the microring lattice with $N = 3$ and $N = 4$ bands.
\begin{figure}[t]
    \centering
    \includegraphics[scale=0.5]{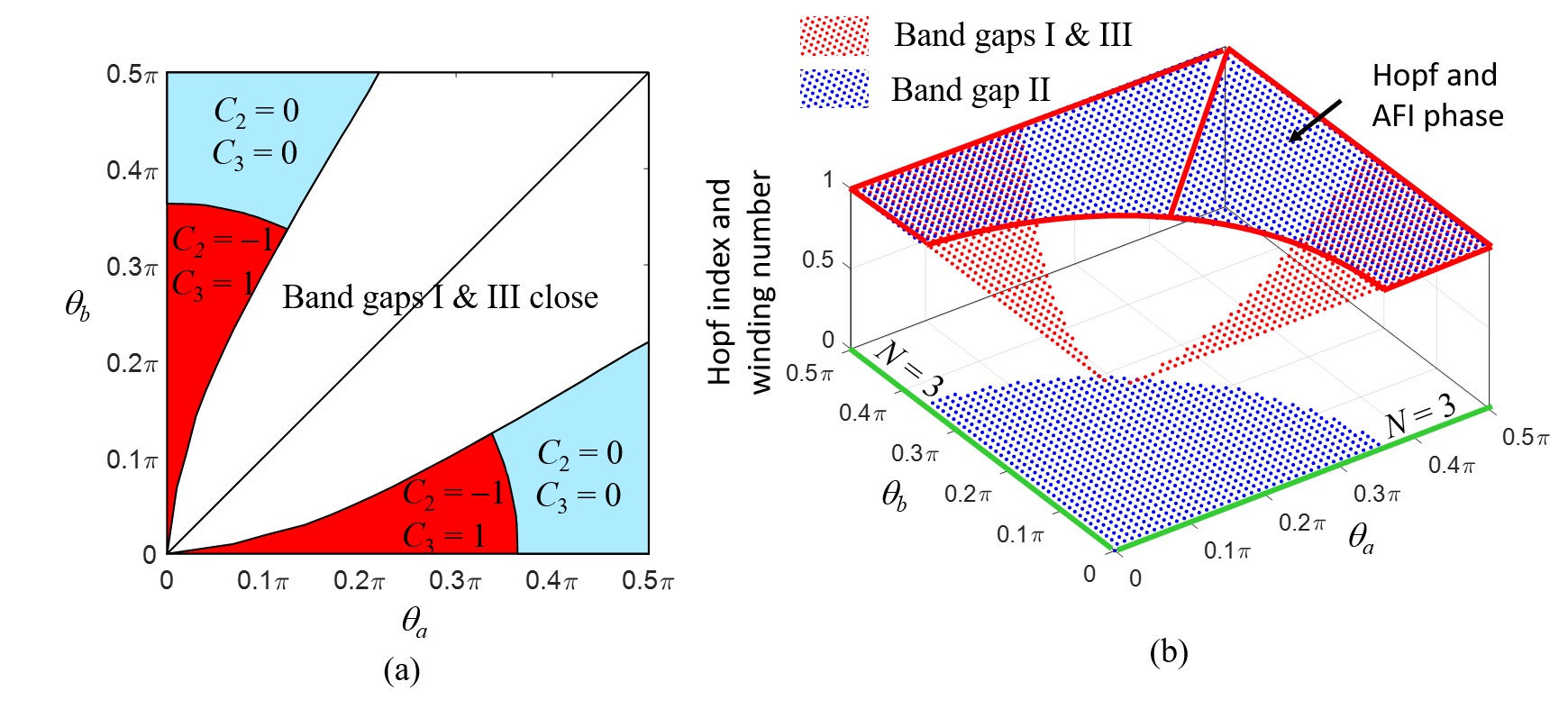}
    \caption{(a) Map of the Chern numbers $C_2$ and $C_3$ of quasi-energy bands $n = 2$ and $n = 3$ of the microring lattice. (b) Map of the $N$-band Hopf index and winding number of each band gap, with the regions of Hopf insulator phase and AFI phase marked by red lines.  The $N = 3$ band lattice is indicated by the green lines along the $\theta_a = 0$ and $\theta_b = 0$ axes.}
    \label{fig:map}
\end{figure}


As mentioned above, there is a correspondence between the Hopf phase and the AFI phase of a Floquet insulator.  We have also calculated the winding number for each band gap of the 2D microring lattice \cite{afzal2018topological}.
The winding number associated with a gap at quasi-energy $\xi$ is computed using the periodized evolution operator as \cite{rudner2013anomalous}
\begin{equation}
\begin{split}
     W[\mathcal{U}_\xi]=&\frac{1}{8\pi^2}\int_0^L dz\int_{BZ} dk_x dk_y \\& \text{Tr}\bigg\{ \mathcal{U}_\xi^{\dagger}\partial_z \mathcal{U}_\xi \left[\mathcal{U}_\xi^{\dagger} \partial_{k_x}\mathcal{U}_\xi,\mathcal{U}_\xi^{\dagger} \partial_{k_y} \mathcal{U}_{\xi}  \right] \bigg\} \label{eq:winding}
    \end{split}
\end{equation}
For the 2D microring lattice, the map of the winding number vs. coupling angles is found to be identical to the Hopf invariant, as shown in Fig. 2(b).  Like the Hopf insulator phase, the AFI phase is defined as having nontrivial winding number when the Chern numbers of all the bands vanish.  Fig. 2(b) shows that the region where the microring lattice exhibits AFI behavior exactly matches with the region for the Hopf phase, thus confirming the correspondence between these two phases for the microring lattice.  This result shows that the $N$-band Hopf index can be used as an alternative invariant to the winding number to classify 2D microring lattices, correctly predicting the existence of topologically protected edge states in each band gap\cite{afzal2020realization}.

The $N$-band Hopf invariant is computed using the $z$-evolved Floquet states while the winding number is computed from the evolution operator.  It is possible to analytically prove the correspondence of these two invariants for a general $N$-band Floquet lattice, which we have included in the Supplementary Materials.  While this correspondence has been shown for 3D static systems with $N$ bands in \cite{lapierre2021n}, our proof provides a direct analytical connection between these two invariants for a 2D+1 Floquet system.  It also shows how the non-Abelian connection matrix $\vec{\hat{A}}$ can also be constructed using states evolving from the standard bases for the computation of the Hopf invariant.

The presence of the Hopf insulator phase is attributed to the nontrivial topological behavior of the entire $N$-band structure which is irreducible to the topology of the individual bands. A constraint specified in \cite{lapierre2021n} for the $N$-band Hopf index to be well defined is that the $N$ bands must be distinct and separated by open gaps; otherwise there may be an integer ambiguity in the Hopf index given by Eq.(\ref{eq:HopfIndex}).  On the other hand, the winding number $W[\mathcal{U}_\xi]$ of the evolution operator does not have such ambiguity.  As a result, the Hopf invariant computed using the $z$-evolved Floquet states, being identical to $W[\mathcal{U}_\xi]$, remains well-defined.  We found that this is the case for the $N = 4$ band microring lattice, which has well-defined Hopf index even although bands $n = 1$ and $n = 4$ are entangled to form a composite band.

\section*{Conclusions}
In summary, 2D microring lattices have previously been shown to emulate 2D+1 Floquet systems which can exhibit AFI phase as classified by the winding number; here we showed that these systems can also be regarded as $N$-band Hopf insulators classified by an alternative invariant called the $N$-band Hopf index.  We also showed by numerical computation and analytical proof that the Hopf invariant is identical to the winding number and established the correspondence between the AFI phase and Hopf insulator phase of the microring lattice. These lattices represent the first realization of photonic Hopf insulators and more generally, can provide a versatile nanophotonic platform for studying non-Abelian topological photonic systems.

\bigskip
\textbf{Data Availability.} Data underlying the results presented in this paper are available in the fig.2(b) of this paper.

\textbf{Funding.} Natural Sciences and Engineering Research Council of Canada.

\textbf{Disclosures.} The authors declare no conflicts of interest.

\bibliographystyle{unsrt}  
\bibliography{references} 
\newpage

\section*{Appendix}

\subsection*{Floquet-Bloch Hamiltonian of 2D square microring lattices} 
Here we derive the Floquet-Bloch Hamiltonian for a 2D square microring lattice shown in Fig. S1(a).  Each unit cell consists of four resonators labeled $A, B, C, D$.
In each microring light propagates in either the counter-clockwise or clockwise direction, which we denote as the $z$ direction.  By cutting each microring at the point indicated by the open circle in Fig. S3(a) and unrolling it into a straight waveguide, we can transform the microring lattice into a 2D array of coupled waveguides as shown in Fig. S3(b) \cite{afzal2018topological}. Each waveguide has length $L$ equal to the microring's circumference, which is divided into a sequence of 4 coupling steps, each step of length $L/4$.  Denoting the fields in the microrings in unit cell $(m,n)$ as $\psi_A^{m,n}, \psi_B^{m,n}, \psi_C^{m,n}, \psi_D^{m,n}$, we can write the coupled mode equations for the evolution of the fields along the waveguide array as
\renewcommand{\thefigure}{S\arabic{figure}}
\begin{figure}[ht!]
    \centering
    \includegraphics[scale=0.35]{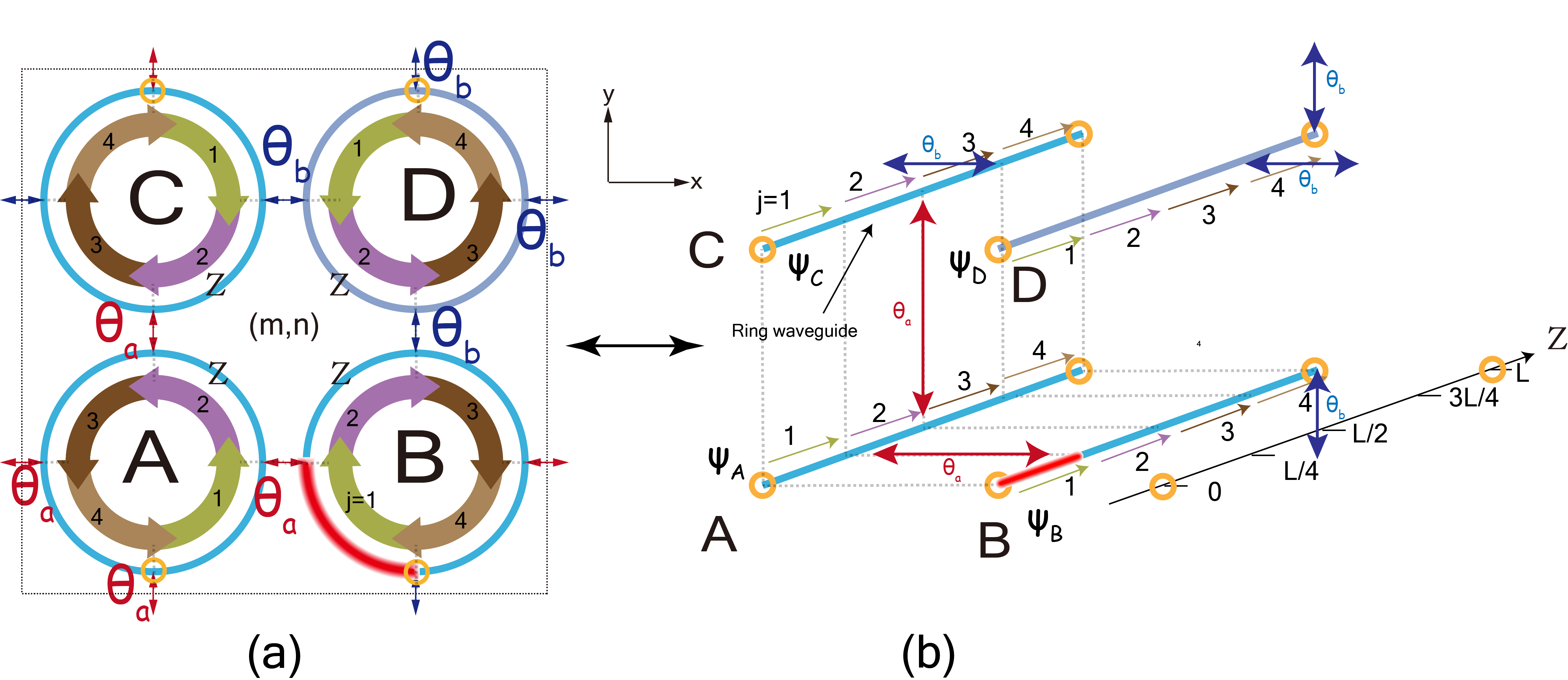}
    \caption{(a) Schematic of a unit cell ($m, n$) of the 2D square microring lattice. (b) Equivalent 2D coupled waveguide array of the unit cell obtained by cutting each microring at the point indicated by the yellow open circle and unrolling it into a straight waveguide.  The propagation of a light over one period $L$ is divided into 4 equal-length segments. An example of light propagating along segment 1 of microring $B$ is shown in red for both the microring lattice and the waveguide array.}
    
\end{figure}

\renewcommand{\theequation}{S\arabic{equation}}

\begin{align} \label{eq:31}
    -i\frac{\partial \psi_A^{m,n}}{\partial z}&= \beta\psi_A^{m,n} +k_a(1)  \psi_B^{m,n}+k_a(2)\psi_C^{m,n}+k_a(3)\psi_B^{m-1,n}+k_a(4) \psi_C^{m,n-1} \nonumber\\
    -i\frac{\partial \psi_B^{m,n}}{\partial z}&=\beta \psi_B^{m,n}+k_a(1)\psi_A^{m,n}+k_b(2)\psi_D^{m,n}   +k_a(3)\psi_A^{m+1,n} + k_b(4)\psi_D^{m,n-1}\nonumber \\
     -i\frac{\partial \psi_C^{m,n}}{\partial z}&= \beta \psi_C^{m,n}+k_b(1)\psi_D^{m,n}+k_a(2)\psi_A^{m,n}   +k_b(3)\psi_D^{m-1,n} + k_a(4)\psi_A^{m,n+1} \nonumber\\
    -i\frac{\partial \psi_D^{m,n}}{\partial z}&=\beta \psi_D^{m,n}+k_b(1)\psi_C^{m,n}+k_b(2)\psi_B^{m,n}   +k_b(3)\psi_C^{m+1,n} +k_b(4)\psi_B^{m,n+1} 
\end{align}
where $\beta$ is the propagation constant of the microring waveguides and the coupling strength $k_{(a,b)}(j)=4\theta_{(a,b)}/L$ in step $j$ and 0 otherwise. 
Applying Bloch's theorem in the $x$ and $y$ directions of the microring lattice, we have
\begin{equation}
    \psi_{A,B,C,D}^{m\pm \mu,n\pm \nu}=\psi_{A,B,C,D}^{m,n}e^{i(\pm \mu k_x \pm \nu k_y)\Lambda}
\end{equation}
where $\Lambda$ is the lattice constant and $\mu$ and $\nu$ are integers. Using the above boundary conditions, we can cast eq.(\ref{eq:31}) in the form of a Schrodinger-like equation for the state vector $\ket{\psi}=[\psi_A^{m,n},\psi_B^{m,n},\psi_C^{m,n},\psi_D^{m,n}]^\mathrm{T}$ of each unit cell,
\begin{equation}
    -i\frac{\partial}{\partial z}|\psi(\vec{k},z)\rangle=[\beta I +H_{FB}(\vec{k},z)]|\psi(\vec{k},z)\rangle
\end{equation}
where $\vec{k}=(k_x,k_y)$ is the crystal momentum vector. The Floquet-Bloch Hamiltonian consists of 4 steps, $H_{FB}=\sum_{j=1}^4 H_j$, with the Hamiltonian in each step explicitly given by
\begin{align} \label{eq:H}
    H_1&=\begin{bmatrix}
    0&k_a& 0& 0 \\
     k_a &0 &0& 0\\
    0& 0& 0& k_b \\
    0& 0& k_b &0\\
    \end{bmatrix}\nonumber \\
    H_2&= \begin{bmatrix}
    0&0&k_a&0\\
    0&0&0&k_b\\
    k_a&0&0&0\\
    0&k_b&0&0\\
    \end{bmatrix}\nonumber \\
       H_3&=\begin{bmatrix}
    0&k_ae^{-ik_x\Lambda}& 0& 0 \\
     k_ae^{ik_x\Lambda} &0 &0& 0\\
    0& 0& 0& k_be^{-ik_x\Lambda} \\
    0& 0& k_be^{ik_x\Lambda} &0\\
    \end{bmatrix}\nonumber \\
     H_4&= \begin{bmatrix}
    0&0&k_ae^{-ik_y\Lambda}&0\\
    0&0&0&k_be^{-ik_y\Lambda}\\
    k_ae^{ik_y\Lambda}&0&0&0\\
    0&k_be^{ik_y\Lambda}&0&0\\
    \end{bmatrix}\nonumber \\
\end{align}
The above expressions can be written in the compact form of Eq.(4) in the main text with $\Lambda=1$. 
The $N=3$ band lattice is obtained by setting either $\theta_a=0$ or $\theta_b=0$. The explicit form of the Floquet Bloch Hamiltonian for the $N = 3$ band lattice can be found in the Supplementary Material of \cite{zimmerling2022broadband}.  
 
 Neglecting the dynamic phase term ($e^{i\beta z}$), the evolution operator of the lattice is given by $\mathcal{U}(\vec{k},z) = \mathcal{T} e^{i\int_0^z H_{FB}(\vec{k},z')dz'}$, where $\mathcal{T}$ is the time-order operator. 
 Since the Hamiltonian $H_j$ in each step $j$ is independent of $z$, we can define the evolution during step $j$ as $\mathcal{U}_j(\vec{k},z) = e^{iH_j(\vec{k})z}$.  The evolution operator over each period can then be explicitly computed as
\begin{equation}
    \mathcal{U}(\vec{k},z)=\Bigg \{ 
    \begin{array}{ll}
    \mathcal{U}_1(\vec{k},z),  & 0\leq z<L/4\\
    \mathcal{U}_2(\vec{k},z-L/4)\mathcal{U}_1(\vec{k},L/4), & L/4\leq z <L/2\\
    \mathcal{U}_3(\vec{k},z-L/2)\mathcal{U}_2(\vec{k},L/4)\mathcal{U}_1(\vec{k},L/4), & L/2\leq z<3L/4\\
    \mathcal{U}_4(\vec{k},z-3L/4)\mathcal{U}_3(\vec{k},L/4)\mathcal{U}_2(\vec{k},L/4)\mathcal{U}_1(\vec{k},L/4), & 3L/4\leq z < L
    \end{array}
\end{equation}

\subsection*{Proof of the correspondence between the $N$-band Hopf index and the winding number}

Here we provide an analytical proof of the correspondence between the Hopf index and the winding number of a 2D $N$-band Floquet insulator with $z$-periodic Hamiltonian $H(k_x,k_y,z)$ (for time-periodic systems, substitute $z \rightarrow t$). 
The $N$-band Hopf invariant is
\begin{equation}
    n_H = \frac{1}{8\pi^2}\int_0^L dz \int_{BZ} Q_3 dk_x dk_y   
    \label{eq:hofp}
\end{equation}
where $L$ is the evolution period and $Q_3$ is the non-Abelian third Chern-Simons form,
\begin{equation}
    Q_3 = \text{Tr}\bigg\{ \vec{ \hat{A}}\cdot \nabla \times \vec{\hat{A}}-\frac{2i}{3} \vec{\hat{A}}\cdot \vec{\hat{A}}\times \vec{\hat{A}}\bigg\} 
    = \epsilon_{jkl}\text{Tr}\bigg\{ \hat{A}_j\partial_k \hat{A}_l-\frac{2i}{3} \hat{A}_j \hat{A}_k \hat{A}_l\bigg\}
    \label{eq_CS}
\end{equation}
with the indices $j, k, l$ running over $k_x, k_y, z$.  We will first construct the connection matrices $\hat{A}_k, k \in \{k_x, k_y, z\}$, using states $|u_m\rangle$ that evolve from a set of $N$ constant orthogonal states $|m\rangle$ at $z = 0$,
\begin{equation}
    |u_m(k_x,k_y,z) \rangle = \mathcal{U}_\xi(k_x,k_y,z) |m\rangle
\end{equation}
where $\mathcal{U}_\xi$ is the periodized evolution operator anchored at band gap quasi-energy $\xi$. For simplicity, we can take $|m\rangle$ to be the $N$-dimensional standard basis vector, with all zero elements except for the $m^{\mathrm{th}}$ element set to 1.  To obtain the connection matrix computed from the $z$-evolved Floquet states in Eq.(8) in the main text, we simply perform the gauge transformation,  $|\Psi_m\rangle = \mathcal{U}_\xi Q_F \mathcal{U}_\xi^\dagger |u_m\rangle$, where $Q_F$ is the eigenvector matrix of the Floquet operator $U_F(k_x,k_y)$.  In the basis of the states $|u_m\rangle$, the $(m,n)$ element of the connection matrix $\hat{A}_k$ is given by $(\hat{A}_k)_{mn}=i\langle u_m | \partial_k u_n\rangle$. The matrix $\hat{A}_k$ can thus be expressed as
\begin{equation}
    \hat{A}_k = i \mathcal{U}_\xi^\dagger \partial_k \mathcal{U}_\xi.
    \label{eq_A}
\end{equation}
Substituting the above expression for $\hat{A}_k$ into Eq.(\ref{eq_CS}), we have
\begin{align}
        Q_3 &= \epsilon_{jkl}\text{Tr}\bigg\{ -\mathcal{U}_\xi^\dagger \partial_j \mathcal{U}_\xi \partial_k(\mathcal{U}_\xi^\dagger\partial_l \mathcal{U}_\xi) - \frac{2}{3}(\mathcal{U}_\xi^\dagger \partial_j \mathcal{U}_\xi)(\mathcal{U}_\xi^\dagger \partial_k \mathcal{U}_\xi)(\mathcal{U}_\xi^\dagger \partial_l \mathcal{U}_\xi) \bigg\} \nonumber \\ 
    &= \epsilon_{jkl}\text{Tr}\bigg\{ -\mathcal{U}_\xi^\dagger \partial_j \mathcal{U}_\xi \partial_k \mathcal{U}_\xi^\dagger \partial_l \mathcal{U}_\xi - (\mathcal{U}_\xi^\dagger \partial_j \mathcal{U}_\xi) \mathcal{U}_\xi^\dagger \partial_k \partial_l \mathcal{U}_\xi \nonumber \\ 
    & -\frac{2}{3}(\mathcal{U}_\xi^\dagger \partial_j \mathcal{U}_\xi)(\mathcal{U}_\xi^\dagger \partial_k \mathcal{U}_\xi)(\mathcal{U}_\xi^\dagger \partial_l \mathcal{U}_\xi) \bigg\}
    \label{eq_CS2}
\end{align}
Since $\hat{A}_k = \hat{A}_k^\dagger$ (due to the orthogonality of $|u_m\rangle$), we have $i \mathcal{U}_\xi^\dagger \partial_k \mathcal{U}_\xi = -i (\partial_k \mathcal{U}_\xi^\dagger) \mathcal{U}_\xi$, which gives $\partial_k \mathcal{U}_\xi^\dagger = -(\mathcal{U}_\xi^\dagger \partial_k \mathcal{U}_\xi) \mathcal{U}_\xi^\dagger$.  Applying this relation to the first term on the right of Eq.(\ref{eq_CS2}), we obtain
\begin{equation}
    Q_3 = \epsilon_{jkl}\mathrm{Tr}\bigg\{ \frac{1}{3}(\mathcal{U}_\xi^\dagger \partial_j \mathcal{U}_\xi)(\mathcal{U}_\xi^\dagger \partial_k \mathcal{U}_\xi)(\mathcal{U}_\xi^\dagger \partial_l \mathcal{U}_\xi) \bigg\} - \epsilon_{jkl}\mathrm{Tr}\bigg\{(\mathcal{U}_\xi^\dagger \partial_j \mathcal{U}_\xi) \mathcal{U}_\xi^\dagger \partial_k \partial_l \mathcal{U}_\xi \bigg \}
\end{equation}
The second term on the right hand side vanishes since $\partial_k$ commutes with $\partial_l$ but $\epsilon_{jkl} = -\epsilon_{jlk}$, yielding
\begin{equation}
    Q_3 = \frac{1}{3}\epsilon_{jkl}\text{Tr}\bigg\{ (\mathcal{U}_\xi^\dagger \partial_j \mathcal{U}_\xi)(\mathcal{U}_\xi^\dagger \partial_k \mathcal{U}_\xi)(\mathcal{U}_\xi^\dagger \partial_l \mathcal{U}_\xi) \bigg\}
    \label{eq_CS3}
\end{equation}
By writing out the terms of the summation over $i, j, k$ and simplifying using the invariance property of the trace under cyclic permutation of the matrices, we obtain
\begin{equation}
    Q_3 = \text{Tr}\bigg\{ (\mathcal{U}_\xi^\dagger \partial_z \mathcal{U}_\xi) \left[(\mathcal{U}_\xi^\dagger \partial_{k_x}\mathcal{U}_\xi)(\mathcal{U}_\xi^\dagger \partial_{k_y} \mathcal{U}_{\xi}) - (\mathcal{U}_\xi^\dagger \partial_{k_y}\mathcal{U}_\xi)(\mathcal{U}_\xi^\dagger \partial_{k_x} \mathcal{U}_{\xi}) \right] \bigg\}
    \label{eq_CS4}
\end{equation}
Substitution of the above expression for $Q_3$ into Eq.(\ref{eq:hofp}) yields the same expression for the winding number, $W[\mathcal{U}_\xi]$, in Eq.(9) in the main text for a 2D+1 Floquet insulator.  

It remains to be shown that when we transform to the basis of the $z$-evolved Floquet states, $|\Psi_m\rangle = \mathcal{U}_\xi Q_F \mathcal{U}_\xi^\dagger |u_m\rangle$, the non-Abelian third Chern-Simons form remains unchanged.  Given the gauge transformation $|\Psi_m\rangle = \mathcal{U}_\xi Q_F \mathcal{U}_\xi^\dagger |u_m\rangle$, the non-Abelian third Chern-Simons forms $P_3$ computed in the bases $|\Psi_m\rangle$ and  $|u_m\rangle$ are related by \cite{ryu2010topological,lapierre2021n}
\begin{equation}
    P_3[\Psi_m] = P_3[u_m] + W[\mathcal{U}_\xi  Q_F \mathcal{U}_\xi^\dagger ]
\end{equation}
The winding number $W[\mathcal{U}_\xi  Q_F \mathcal{U}_\xi^\dagger ]$ can be expressed as \cite{weinberg2012classical}
\begin{equation}
    W[\mathcal{U}_\xi  Q_F \mathcal{U}_\xi^\dagger ]=W[\mathcal{U}_\xi] + W[ Q_F] + W[ \mathcal{U}_\xi^\dagger ]
\end{equation}
Since $\mathcal{U}_\xi\mathcal{U}_\xi^\dagger=I$, we have $W[\mathcal{U}_\xi]+W[\mathcal{U}_\xi^\dagger]=0$.  Furthermore $W[Q_F] = 0$ since $Q_F$ is independent of $z$.  Thus $W[\mathcal{U}_\xi  Q_F \mathcal{U}_\xi^\dagger ] = 0$ so that $P_3[\Psi_m] = P_3[u_m]$. Thus the Hopf index remains unchanged if we construct the connection matrix $\vec{\hat{A}}$ from the $z$-evolved Floquet states instead of the states $|u_m\rangle$.

\end{document}